\documentclass{aa}  

\usepackage{graphicx}
\usepackage{txfonts}

\usepackage[]{hyperref}
\newcommand{\msun}{\mathrm{M_\odot}}	% solar mass

\begin{document}

   \title{Automated galaxy sizes in Euclid images using the Segment Anything Model}

   %\subtitle{I. Overviewing the $\kappa$-mechanism}

    \author{J. Vega-Ferrero\inst{1,2}\thanks{E-mail: astrovega@gmail.com}
      \and F. Buitrago\inst{1,3}
      \and J. Fernández-Iglesias\inst{4}
      \and S. Raji\inst{1}
      \and B. Sahelices\inst{4}
      \and H. Domínguez Sánchez\inst{2,5}}
    
    \institute{%
      \inst{1} Departamento de Física Teórica, Atómica y Óptica, Universidad de Valladolid, 47011 Valladolid, Spain\\
      \inst{2} Centro de Estudios de Física del Cosmos de Aragón (CEFCA), Plaza de San Juan, 1, E-44001 Teruel, Spain\\
      \inst{3} Instituto de Astrofísica e Ciências do Espaço, Universidade de Lisboa, OAL, Tapada da Ajuda, PT1349-018 Lisbon, Portugal\\
      \inst{4} GIR GCME. Departamento de Informática, Universidad de Valladolid, 47011 Valladolid, Spain\\
      \inst{5} Instituto de Física de Cantabria (CSIC-UC), Avda. Los Castros s/n, E-39005 Santander, Spain
    }
   %\date{Received September 15, 1996; accepted March 16, 1997}
   \date{Received XXX; accepted YYY}

% \abstract{}{}{}{}{} 
% 5 {} token are mandatory
 
  \abstract
   {Stellar disk truncations, also referred to as galaxy edges, are key indicators of galactic size, determined by the radial location of the gas density threshold for star formation. This threshold essentially marks the boundary of the luminous matter in a galaxy. Accurately measuring galaxy sizes for millions of galaxies is essential for understanding the physical processes driving galaxy evolution over cosmic time.}
   {We aim to explore the potential of the Segment Anything Model (SAM), a foundation model designed for image segmentation, to automatically identify disk truncations in galaxy images. With the Euclid Wide Survey poised to deliver vast datasets, our goal is to assess SAM’s capability to measure galaxy sizes in a fully automated manner.}
   {SAM was applied to a labeled dataset of 1,047 disk-like galaxies with $M_* > 10^{10} \msun$ at redshifts up to $z \sim 1$, sourced from the Hubble Space Telescope (HST) CANDELS fields. We ‘euclidized’ the HST galaxy images by creating composite RGB images, using the F160W (H-band), F125W (J-band), and F814W + F606W (I-band + V-band) HST filters, respectively. Using these processed images as input for SAM, we retrieved various truncation masks for each galaxy image under different configurations of the input data.}
   {We find excellent agreement between the galaxy sizes identified by SAM and those measured manually (i.e., by using the radial positions of the stellar disk edges in galaxy light profiles), with an average deviation of approximately $3\%$. This error reduces to about $1\%$ when excluding problematic cases.}
    {Our results highlight the strong potential of SAM for detecting disk truncations and measuring galaxy sizes across large datasets in an automated way. SAM performs well without requiring extensive image preprocessing, labeled training datasets for truncations (used only for validation), fine-tuning, or additional domain-specific adaptations such as transfer learning.}


   \maketitle
%
%-------------------------------------------------------------------

\section{Introduction}

In astrophysics, there are indeed several methods for measuring a galaxy's size, but all come with significant challenges due to the diffuse and often irregular nature of galaxies, which don't have well-defined boundaries. Various strategies have been explored over the past hundred years, with \cite{Chamba20_hist} providing a comprehensive overview. The effective radius ($r_e$), defined as the semi-major axis of an ellipse that encompasses half of a galaxy's light, has emerged as the predominant measure of galaxy size in recent times. Its widespread adoption is primarily due to its reliability across various signal-to-noise ratios (S/N) and exposure times, as well as its integration with the parametric fitting of galaxy surface brightness, specifically through the \cite{Sersic1968} functions. Nonetheless, the choice of using the half-light radius as a size metric is somewhat arbitrary and deeply linked to a galaxy's concentration, a point elaborated upon in the introductory section of \cite{Trujillo20}.

\cite{Trujillo20} introduces the R1 (the isomass contour at $1 \mathrm{M_\odot/pc^2}$) as a size indicator grounded in physical principles. The current galaxy formation model suggests an initial phase of star formation within a massive galaxy, followed by the accretion of satellite galaxies. This initial phase's end should leave a mark on the outer regions of the galaxy, where the gas density necessary for star formation becomes unattainable, as argued by \cite{Schaye04}. This transition from gas to stellar density, albeit at a low efficiency rate, explains why R1 marks a significant shift in galaxy attributes (such as surface brightness, color, and mass distribution) for galaxies similar to the Milky Way in our local Universe \citep{Martinez-Lombilla19,Diaz-Garcia22}. However, the applicability of the $1~\mathrm{M_\odot/pc^2}$ threshold varies, depending mainly on the galaxy's stellar mass or the specific conditions of the star-forming event. \cite{Chamba2022} and \cite{Buitrago2024} have investigated these abrupt shifts in galaxy outer profiles across different stellar masses and redshifts, revealing the evolution of these features. The observed reduction in the scatter of the mass-size relationship by 2-2.5 times, endorse this method as a physically-based approach to defining galaxy dimensions. Interestingly, these low surface brightness (LSB) features have been known since the late 70s \citep{vanderKruit1979}, but the shallow nature of most galaxy survey images prevented a better understanding of this topic.

Galaxy truncations were first spotted in edge-on galaxy disks, which maintained their sizes despite prolonged exposure times \citep{vanderKruit1981a,vanderKruit1981b}. Importantly, this observation does not contradict the diffuse nature of galaxies, as the presence of a stellar halo or stellar migration ensures stars and light extend beyond these truncations. Previously, these truncations were not deemed effective for size measurement due to their appearance at  LSB levels, $\mu > 26-27~\mathrm{mag/arcsec^2}$ in $10\times10$ arcsec apertures \citep{Martin-Navarro12,Martin-Navarro14, Trujillo16}. However, according to \cite{Borlaff22}, this perspective is set to change with the advent of future telescopes capable of detecting extremely LSB levels ($\mu > 28-30~\mathrm{mag/arcsec^2}$ (such as LSST, Roman Space Telescope and ARRAKIHS mission), leading to new insights into the LSB domain of galaxy formation and evolution \citep{Duc15,Mihos19}. Yet, the challenge ahead lies not only in constructing advanced telescopes and instruments, but also in developing sophisticated software capable of accurately analyzing the vast number of celestial objects that these instruments will uncover.

Determining galaxy truncations is an object segmentation task within the field of computer vision, especially when approached from the perspective of analyzing astronomical images to identify and delineate specific features of galaxies. Object segmentation in computer vision involves dividing an image into segments or regions that are significant and relevant to the task at hand, often to isolate objects of interest from the background or to distinguish between different objects within the image. Galaxy outskirts display sudden drops in their surface brightness and mass profiles, which allows the truncation detection to be clearly addressed as a pattern recognition problem. Advanced techniques involving pattern recognition and machine learning (ML) can be trained to automatically identify patterns and features for: the classification of galaxy images \citep[][to name a few examples]{Dieleman2015,Huertas-Company15,dominguez-sanchez18,Hausen20,Walmsley2020,vega-ferrero21,Walmsley2022,Vega-Ferrero2024}; object detection and segmentation using different deep learning (DL) models \citep{Burke2019,Gonzalez2018,Paillassa2020,Farias2020,Tanoglidis2022} and U-Nets \citep{Hausen20,Boucaud2020,Bretonniere2021,Fernandez2024}. In particular, \citet{Fernandez2024} presents a novel U-Net approach to automatically determine the position of galaxy truncations for large datasets of galaxy images. Their findings suggest that similar performances than humans could be routinely achieved by combining the output of several neural networks (NNs) using ensemble learning, provided a large enough training sample of previously segmented galaxy images of the same data domain is available.

Indeed, previous object segmentation methods typically require large amounts of annotated data for training, making them difficult to apply to new or unseen objects or tasks. The Segment Anything Model (SAM\footnote{\url{https://segment-anything.com/}}, \citealt{Kirillov2023}) addresses this challenge by leveraging a large synthetic dataset, trained with a prompt-based approach, enabling it to generalize to unseen objects and tasks without additional training. As an artificial intelligence (AI) foundational model \citep{bommasani2021}, SAM can be adapted to various downstream tasks, breaking traditional boundaries of segmentation tasks. SAM represents a significant advancement in the field of AI, particularly in computer vision and image processing, being able to segment and identify objects within any given image, regardless of the context or the complexity of the scene, making it a valuable tool for a wide range of applications \citep{Zhang2023}, including astrophysics. For instance, by utilizing natural language prompts to specify the desired objects, astronomers could effortlessly segment specific components of complex astronomical images, and classify them based on their morphological \citep{Tanoglidis2024}, photometric and spectral properties; by providing a prompt to identify objects that deviate from the norm, astronomers could automate the process of identifying anomalies as potential sources of interest in astrophysical research projects; and by training the model on large datasets of labeled astronomical images, researchers could develop automated systems for classifying objects based on their morphological, photometric and spectral properties \citep{Parker2024}. In this study, we propose and validate the use of SAM, an AI foundational model for segmentation, to identify stellar disk truncations in Euclid-like images automatically. Subsequently, we infer physically-motivated galaxy sizes using the truncations obtained with SAM from a set of approximately one thousand HST galaxies previously studied in \citet{Buitrago2024}. We demonstrate the enormous potential of SAM in estimating galaxy sizes for large samples rapidly and accurately, without the need for training the model with previously known truncations or performing any transfer learning or domain adaptation step.

The paper is structured as follows: \autoref{sec:dataset} presents the dataset of galaxy images used; \autoref{sec:truncations} describes the methodology applied to infer galaxy truncations; \autoref{sec:results} details our results; and \autoref{sec:conclusions} reports our main conclusions. Throughout the paper, our assumed cosmology is $\Omega_m=0.3$, $\Omega_{\Lambda}=0.7$ and $H_0=70 \mathrm{km~s^{-1}~Mpc^{-1}}$, along with a stellar initial mass function from \citep{Chabrier03}.

%%%%%%%%%%%%%%%%%%%%%%%%%%%%%%%%%%%%%%%%%%%%%%%%%%%%%%%%%%%%%%%%%%%%%%%%%%%%%%%%%%%%%%%%%%%%%%%%%%%%
%%%%%%%%%%%%%%%%%%%%%%%%%%%%%%%%%%%%%%%%%%%%%%%%%%%%%%%%%%%%%%%%%%%%%%%%%%%%%%%%%%%%%%%%%%%%%%%%%%%%
%%%%%%%%%%%%%%%%%%%%%%%%%%%%%%%%%%%%%%%%%%%%%%%%%%%%%%%%%%%%%%%%%%%%%%%%%%%%%%%%%%%%%%%%%%%%%%%%%%%%

\section{Dataset}
\label{sec:dataset}

For our analysis, we utilized the sample of 1047 galaxies described in \citet[][hereafter BT24]{Buitrago2024}. This dataset consists of a redshift- and a mass-selected sample of disk-dominated galaxies. We use the most update set of images (v1.0) in the CANDELS survey\footnote{\url{http://arcoiris.ucolick.org/candels/}} \citep{Grogin11,Koekemoer11} and in the GOODS-South field from the Hubble Legacy Field\footnote{\url{https://archive.stsci.edu/prepds/hlf/}} \citep[HLF v2.0;][]{Illingworth16}. Galaxies have been selected according to their stellar masses ($M_* > 10^{10} M_\odot$) and spectroscopic redshifts ($z_{spec} < 1.1$). This target selections is based on the CANDELS public catalogs\footnote{\url{https://archive.stsci.edu/prepds/candels/}} \citep{Santini15,Stefanon17,Nayyeri17,Barro19}, on the high-quality spectroscopic redshifts from the LEGA-C DR2 redshifts \citep{vanderWel16,Straatman18} and the DR3 data \citep{vanderWel21}, the ZCOSMOS Final Data Release \citep{Lilly09}, and hCOSMOS \citep{Damjanov19}. Galaxies in this dataset are selected to be disk-dominated according to the ML morphological classifications derived by \cite{Huertas-Company15}, split into three categories: pure discs (DISK), disks with central spheroids (DISKSPH), and irregular discs (DISKIRR). In BT24, the authors derive the radial location of the gas density threshold for star formation as a size indicator for the whole dataset using a combination of surface brightness, color, and stellar mass density profiles. We use these stellar truncations and truncation masks as a reference dataset for evaluating our results.

%%%%%%%%%%%%%%%%%%%%%%%%%%%%%%%%%%%%%%%%%%%%%%%%%%%%%%%%%%%%%%%%%%%%%%%%%%%%%%%%%%%%%%%%%%%%%%%%%%%%
%%%%%%%%%%%%%%%%%%%%%%%%%%%%%%%%%%%%%%%%%%%%%%%%%%%%%%%%%%%%%%%%%%%%%%%%%%%%%%%%%%%%%%%%%%%%%%%%%%%%
%%%%%%%%%%%%%%%%%%%%%%%%%%%%%%%%%%%%%%%%%%%%%%%%%%%%%%%%%%%%%%%%%%%%%%%%%%%%%%%%%%%%%%%%%%%%%%%%%%%%

\section{Stellar disk truncations with SAM}
\label{sec:truncations}

\begin{figure*}
\centering
    \begin{minipage}[b]{0.5\columnwidth}
        \centering
        \textbf{($q \times s = 50$, $c = 1$)}\\
        \includegraphics[width=\textwidth]{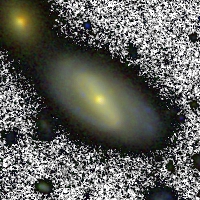}
    \end{minipage}
    \begin{minipage}[b]{0.5\columnwidth}
        \centering
        \textbf{($q \times s = 50$, $c = 5$)}\\
        \includegraphics[width=\textwidth]{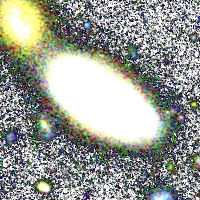}
    \end{minipage}
    \begin{minipage}[b]{0.5\columnwidth}
        \centering
        \textbf{($q \times s = 400$, $c = 1$)}\\
        \includegraphics[width=\textwidth]{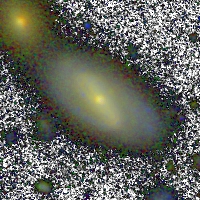}
    \end{minipage}
    \begin{minipage}[b]{0.5\columnwidth}
        \centering
        \textbf{($q \times s = 400$, $c = 5$)}\\
        \includegraphics[width=\textwidth]{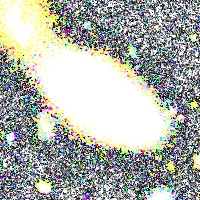}
    \end{minipage}

    \caption{Four RGB images of the DISK galaxy (ID = 1298 COSMOS, $M_* = 1.4 \times 10^{10} \msun$ at $z=0.15$). The RGB channels correspond to the H, J and I+V HST filters, respectively. Images are produced with the Gnuastro script \texttt{astscript-color-faint-gray} for $q \times s = 50$ (first and second panels) and $q \times s = 400$ (third and fourth panels), and for $c = 1$ (first and third panels) and $c = 5$ (second and fourth panels). Stamps have a field of view of $12 \times 12$ arcsec$^2$ ($200 \times 200$ px$^2$).}
    \label{fig:disk_1298_cosmos}
\end{figure*}

By design, SAM has an enormous potential to identify features within astronomical images without the need for extensive preprocessing or a labeled dataset. SAM's architecture includes three main components: an image encoder, a prompt encoder, and a mask decoder. The image encoder is based on a Vision Transformer (ViT) that has been pre-trained to handle high-resolution inputs efficiently. The prompt encoder can process various types of prompts, including sparse (points, boxes, text) and dense (masks), facilitating flexible and accurate segmentation outcomes.  This design allows SAM to generate valid segmentation masks from ambiguous prompts, showcasing its adaptability and potential for application in a wide range of scenarios. The model has been trained and evaluated on a large dataset, SA-1B, which contains over 11 million diverse, high-resolution images and 1.1 billion high-quality segmentation masks. Unlike traditional segmentation models that require extensive pre-labeled datasets for specific objects, SAM utilizes a more generalized approach, allowing it to learn from a broader set of image data and apply its learning to segment objects in previously unseen categories (i.e., zero-shot learning).  Besides, contrarily to the typical design of neural networks, SAM's input images are not necessarily of a fixed size and, therefore, input galaxy images might cover a different field of view, for instance, proportional to a particular size estimate such as the effective radius, while keeping the original pixel size.

A segmentation mask in computer vision is a binary or multi-class representation of an image where each pixel is labeled to indicate whether it belongs to a specific object or region of interest. In this study, segmentation masks delineate the stellar disk truncations in galaxy images, distinguishing pixels that belong to galaxies from those associated with the background. SAM produces segmentation masks at a pixel level without assuming any underlying model, and, therefore, SAM truncations are not (a priori) of a particular shape. They can be as irregular as the edges of the sources to be segmented, contrary to the methodology described in BT24, in which truncations are elliptical by definition. Although not strictly forced by design, truncations derived by \citet{Fernandez2024} using U-Nets exhibit also this elliptical shape as their model is trained with the elliptical truncations derived by BT24. This is also a capability that makes SAM an extremely interesting model not only for segmenting stellar disk truncations in galaxy images but also for source detection and characterization of regions with different properties within the same galaxy, especially irregular ones.
%%%%%%%%%%%%%%%%%%%%%%%%%%%%%%%%%%%%%%%%%%%%%%%%%%%%%%%%%%%%%%%%%%%%%%%%%%%%%%%%%%%%%%%%%%%%%%%%%%%%
%%%%%%%%%%%%%%%%%%%%%%%%%%%%%%%%%%%%%%%%%%%%%%%%%%%%%%%%%%%%%%%%%%%%%%%%%%%%%%%%%%%%%%%%%%%%%%%%%%%%

\subsection{Image preprocessing}
\label{sec:data_preprocessing}

For the BT24 dataset, we produced squared stamps with a field of view of 12$\times$12 arcsec$^2$ (i.e., 200$\times$200 px$^2$, with a pixel scale of $0.06$ px/arcsec) in the F606W (V-band) and the F814W (I-band) from the ACS camera, and the F125W (J-band) and the F160W (H-band) from the WFC3 camera. Although, as stated above, the size of the input images could be variable, we choose not to make the cutouts proportional to their size ---which is common in other classification tasks, because size is precisely the parameter we want to infer. We note that all but 11 galaxies in our dataset have apparent sizes that perfectly fit into the field of view of our images, ensuring that no light is lost in our cutouts. On the other hand, the smallest galaxies have apparent sizes above 20 pixels, making it possible to infer a reliable segmentation. Our aim is to test the performance on Euclid data to be ready to apply our methodology for DR1. Therefore, Until Euclid's wide-field survey data is released, we produced mock Euclid (`euclidized') galaxy images by generating composite RGB images using H, J, and I+V HST filters, respectively. We use a new program, \texttt{astscript-color-faint-gray}, firstly introduced in Gnuastro\footnote{\url{https://www.gnu.org/software/gnuastro/}} v0.22 and described in \cite{Infante-Sainz2024}, to generate color images that accurately represent the full dynamic range of the galaxy stamps. It employs an $asinh$ transformation to assign RGB colors to bright pixels (as described in \citealt{Lupton2004}), while the faint ones are shown in an inverse gray-scale, as follows: $f(I) = \frac{asinh(q \times s \times I)}{q}$, where $I$ is the value in each pixel of the image, $s$ is the "stretch" parameter ($s \rightarrow 0$, equivalent to linear stretch), and $q$ is the "bright threshold" that controls the coloring of brighter features. When combining the three images in the H, J, and I+V filters into a composite RGB image, the relevant quantity is the product $q \times s$. Additionally, the "contrast" parameter (denoted as $c$) varies the contrast enhancement of the image. The $asinh$ transformation enhances the faint structures in the outskirts of the central galaxy in each image, while the product $q \times s$ and $c$ strengthen the contrast between the central galaxy and the rest of the image. In this study, the rest of the parameters in the script are not modified to simplify the preprocessing phase of the galaxy images in the dataset.
%%%%%%%%%%%%%%%%%%%%%%%%%%%%%%%%%%%%%%%%%%%%%%%%%%%%%%%%%%%%%%%%%%%%%%%%%%%%%%%%%%%%%%%%%%%%%%%%%%%%
%%%%%%%%%%%%%%%%%%%%%%%%%%%%%%%%%%%%%%%%%%%%%%%%%%%%%%%%%%%%%%%%%%%%%%%%%%%%%%%%%%%%%%%%%%%%%%%%%%%%

\subsection{SAM truncations}
\label{sec:sam_truncations}

Given that we know the position of the object in the image to be segmented (always centered in the image), we call SAM in the informed segmentation mode. In this mode, the user can indicate points, boxes, or regions in the image to be explicitly segmented or avoided. In our dataset, the galaxies are centered in the image so we inform SAM to generate segmented maps including the central pixel of each image. For each galaxy image, SAM infers three different segmented regions with different scores (assigned automatically by SAM) in a map of the same size as the input images. 

%%%%%%%%%%%%%%%%%%%%%%%%%%%%%%%%%%%%%%%%%%%%%%%%%%%%%%%%%%%%%%%%%%%%%%%%%%%%%%%%%%%%%%%%%%%%%%%%%%%%

\subsubsection{Selection of optimal input configurations}
\label{sec:selection}

\begin{figure}
\centering
	\includegraphics[width=\columnwidth]{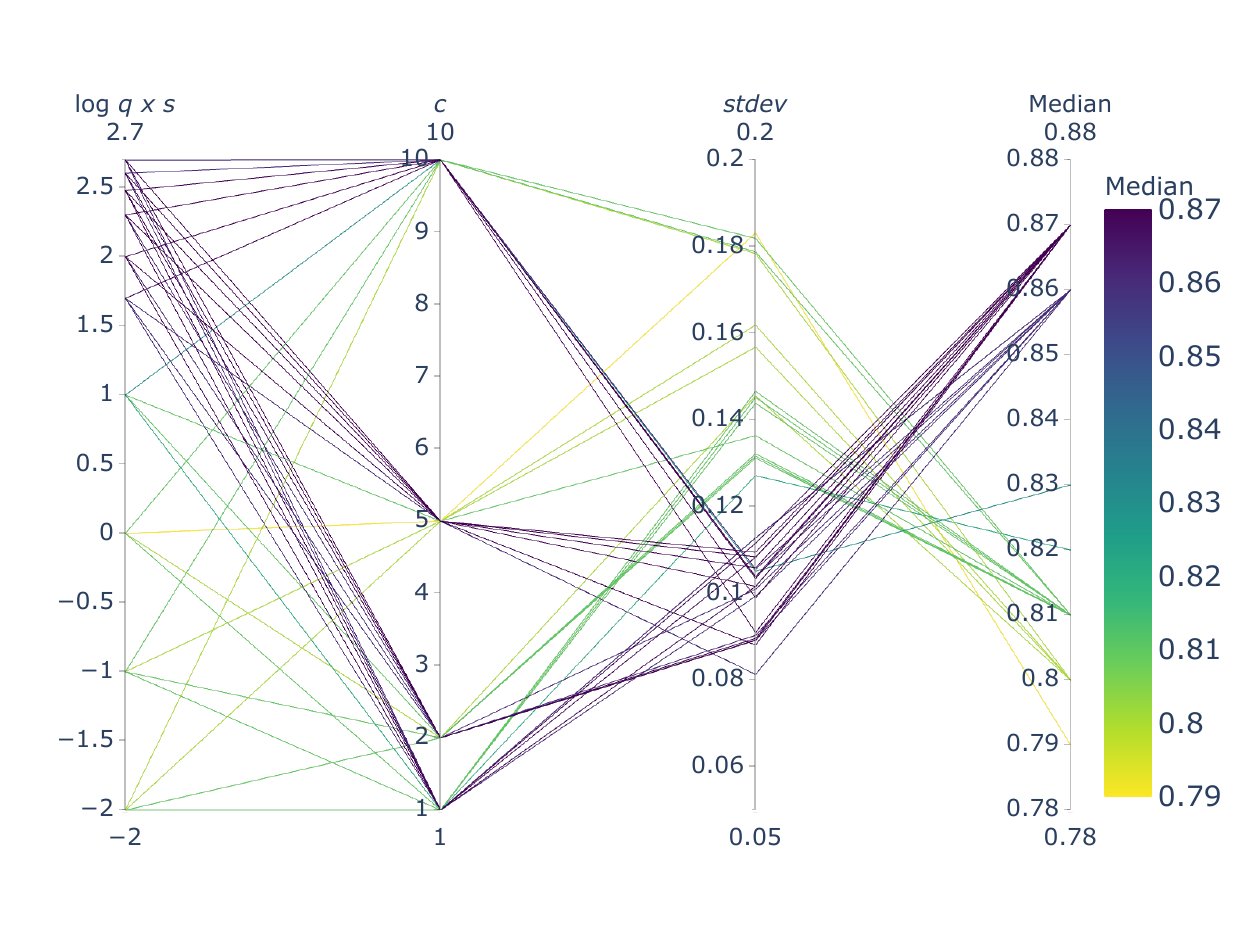}
	
    \caption{Parallel coordinates plot for the 40 input configurations combining $q \times s$ (in log scale, first coordinate), $c$ (second coordinate), standard deviation of the $F_1$ score (third coordinate), and median values of the $F_1$ score (fourth coordinate). Each configuration is color-coded according to the median $F_1$ score. All the configurations with $\log (q \times s) \gtrsim 1.7$ ($q \times s \geq 50$) (independently of the contrast value, $c$) lead to the largest median values of $F_1$ score with the smallest values of the standard deviation of the $F_1$ score.}
    \label{fig:parallel_coords}
\end{figure}

To check the robustness of SAM in retrieving stellar disk truncations we apply it to the whole dataset of RGB images for a large range of combinations of $q \times s$ and $c$. We generate RGB images for $q \times s = (0.01, 0.1, 1, 10, 50, 100, 200, 300, 400, 500)$ and $c = (1, 2, 5, 10)$. In total, we apply SAM over the whole dataset for 40 different input configurations of the input images. In \autoref{fig:disk_1298_cosmos}, we show examples of RGB images of a DISK galaxy $\mathrm{ID}=1298$ in the COSMOS field for $q \times s = (50, 400)$ and $c = (1,5)$.

We compared our results with the truncations presented in BT24 based on the segmentation maps provided in both cases. We should note that our intention is not to consider the findings of BT24 as the definitive ground truth for disk truncations. Instead, our comparison in this section aims to highlight which parameters in the preprocessing phase may be less optimal for the retrieval of reliable truncation signatures. To do so we consider \textit{positive} pixels in the images belonging to the truncation masks from BT24 or SAM predictions, while \textit{negative} pixels are those that fall outside the truncation masks. We use three typical metrics for estimating the model performance, precision (\textit{Prec}), recall (\textit{Rec}) and $F_1$ score (also known as Dice coefficient), according to the \textit{positive}/\textit{negative} pixels in the truncation and SAM predicted masks. We use three typical metrics for estimating the model performance: precision (\textit{Prec}), recall (\textit{Rec}), and $F_1$ score (also known as Dice coefficient), according to the \textit{positive}/\textit{negative} pixels in the truncation and SAM predicted masks. Precision (\textit{Prec}) is defined as the ratio of \textit{true positive} (\textit{TP}) pixels to the sum of \textit{true positive} (\textit{TP}) and \textit{false positive} (\textit{FP}) pixels (\textit{Prec} = $\frac{TP}{TP + FP}$). Recall (\textit{Rec}) is the ratio of \textit{true positive} (\textit{TP}) pixels to the sum of \textit{true positive} (\textit{TP}) and \textit{false negative} (\textit{FN}) pixels (\textit{Rec} = $\frac{TP}{TP + FN}$). The $F_1$ score is the harmonic mean of precision and recall ($F_1 = \frac{2 \cdot \textit{Prec} \cdot \textit{Rec}}{\textit{Prec} + \textit{Rec}}$). Precision can be thought of as purity since it measures the accuracy of positive predictions, while recall is equivalent to completeness as it assesses how thoroughly all actual positives are captured. Then, we compute the $F_1$ score for all the images considering the 40 input configurations. For the three different masks that SAM infers for each image and configuration, we select the one with the highest $F_1$ score. The median value of the $F_1$ score for the 40 configurations ranges from 0.79 to 0.87. As shown in \autoref{fig:parallel_coords}, The configurations with the largest $F_1$ scores, always $F_1 \geq 0.86$, are those with the largest values of $q \times s > 50$, independently of the $c$ value. Hereafter, we discuss the results obtained for these 24 configurations with $q \times s = (50, 100, 200, 300, 400, 500)$ and $c = (1, 2, 5, 10)$.

\begin{figure*}
\centering
	\includegraphics[width=\columnwidth]{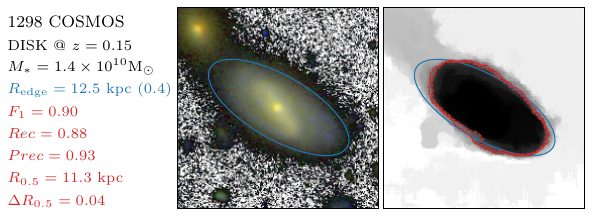}
	\includegraphics[width=\columnwidth]{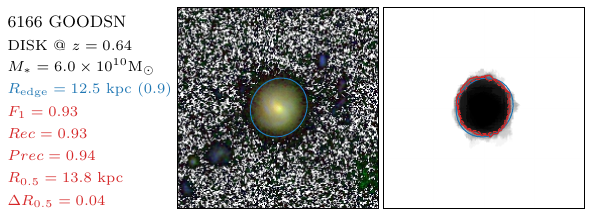}
	
    \caption{SAM truncations for two example galaxies: $\mathrm{ID}=1298$ in the COSMOS field (left-hand panel) and $\mathrm{ID}=6166$ in the GOODSN field (right-hand panel). For each galaxy, we show an RGB image (left-hand panel) of the galaxy and the averaged segmentation map inferred with SAM (right-hand panel). The averaged segmentation map shows the SAM truncations (in gray) for different configurations of the input image, where darker regions indicate stronger agreement between different truncation estimates. The solid red contours depict the SAM truncation based on the majority voting criteria (above a 0.5 threshold), while the dashed red contours correspond to the regions above a 0.7 probability threshold. The blue ellipse corresponds to the truncation derived by BT24 parameterized by the semi-major axis (denoted as $R_{edge}$) and the axis ratio $b/a$ (shown between brackets after the $R_{edge}$ value). We indicate in red the different metrics ($F_1$, $Rec$ and $Prec$) when comparing the SAM and the BT24 truncation estimates, the size of the truncation ($R_{0.5}$), and  the relative error in the size of the truncation and the contour level at 0.7 probability, denoted as $\Delta R_{0.5}$.}
    \label{fig:sam_truncations}
\end{figure*}

%%%%%%%%%%%%%%%%%%%%%%%%%%%%%%%%%%%%%%%%%%%%%%%%%%%%%%%%%%%%%%%%%%%%%%%%%%%%%%%%%%%%%%%%%%%%%%%%%%%%

\subsubsection{Averaged segmentation maps}
\label{sec:averaged_maps}

We take advantage of the segmentations obtained with the different configurations to construct a robust segmentation mask and to quantify uncertain regions. We produce an averaged map with the 24 different configurations selected at the end of \autoref{sec:selection} and the three segmentation maps per each configuration that SAM infers. A total of 72 ($24 \times 3$) segmentation maps are combined to obtain the average one. Segmentation maps are in binary format: zeros for regions outside the truncation, and ones for regions inside it. Therefore, the averaged segmentation map ranges from 0 to 1 values per pixel: 0 means none of the segmentation maps include that pixel as part of the truncations; 1 means all the segmentation maps consider that pixel as part of the truncation; and intermediate values between 0 and 1, reflect different levels of agreement between the different segmentation maps per each configuration. By producing these averaged maps we can determine whether the models converge to a similar solution for the truncation or not. For instance, images containing not only the central galaxy but other sources may lead to significantly different truncations for configurations with different $q \times s$ and $c$ values. High contrast values could blend two distinct galaxies into one single object, producing different truncation masks, as shown in \autoref{fig:disk_1298_cosmos}. Contrarily, in some cases, large $q \times s$ and $c$ values increase the contrast between the faint outskirts of a galaxy and its surrounding background favoring a reliable determination of its truncation using SAM.

We define the best truncation from the averaged segmented maps by applying a majority voting criteria. In other words, we compute the SAM truncation as the contour encompassing the regions with an averaged value above 0.5 (i.e., $>50\%$ of the segmentation maps agree). We also infer the size of the truncation, denoted as $R_{0.5}$ as the maximum distance from the galaxy's center to the SAM truncation. Defining the size of the SAM truncation in this way allows us to reliably compare with the semi-major axis of the elliptical truncation presented in BT24, as shown in \autoref{sec:sizes}. To estimate the robustness of the SAM predictions, we define the relative error in the measurement of the size of the truncation following: $\Delta R_{0.5} = (R_{0.5} - R_{0.7}) / R_{0.5}$, where $R_{0.7}$ is the size of the truncation enclosed by the contour at the 0.7 probability threshold. By definition, $R_{0.7} \leq R_{0.5}$, and, therefore, $\Delta R_{0.5}$ range from 0 to 1 values. By looking at the values of $\Delta R_{0.5}$, we can easily determine whether the different contour levels in the averaged segmentation maps are in agreement or not and, therefore, estimate the robustness of the truncations obtained. Large values of $\Delta R_{0.5}$ indicate that there is a significant difference in the measured size of the truncation (applying a majority voting criteria) and the area enclosed by a higher probability (e.g., 0.7) contour level. This is mainly due to the presence of other sources (i.e., companions, artifacts, or star spikes) close to the truncation inferred by SAM. In summary, $\Delta R_{0.5}$ is certainly a way to determine whether the truncation and its size ($R_{0.5}$) are affected by a contaminant that has been considered as part of the truncation inferred by SAM. Nevertheless, it will be possible to identify galaxies with close companions (and even mask them) from the source catalogs that will be published along with the Euclid images and, therefore, exclude galaxies with less reliable SAM truncations and sizes in subsequent analysis.

In \autoref{fig:sam_truncations}, we show the averaged segmentation maps for two example galaxies: DISK galaxy $\mathrm{ID}=1298$ in the COSMOS field (same as in \autoref{fig:disk_1298_cosmos}) and $\mathrm{ID}=6166$ in the GOODSN field. Although $\mathrm{ID}=1298$ in the COSMOS is not a trivial case due to the presence of a bright companion galaxy in the top-left corner of the image, by looking at the different metrics (e.g., $F_1 = 0.90$), there is a good agreement between truncations derived by BT24 and the SAM truncation. The fraction of pixels (with respect to the total field of view) with low probability values (e.g., below 0.5, light gray pixels) in the averaged segmentation map is high, which indicates that some configurations of the input image include the neighbor galaxy as part of the truncation of the central galaxy. Nevertheless, the majority voting criteria does not include those regions within the truncation (i.e., the 0.5 contour level) favoring a high value of the $Prec = 0.93$. Moreover, a large fraction of the truncation derived by BT24 is also included into the SAM truncation, as reflected by $Rec = 0.88$. In the case of $\mathrm{ID}=6166$ in the GOODSN field, there is a clear agreement between the different SAM predictions for the whole set of configurations. The majority of the predictions clearly depict the edge of the luminous matter in this particular galaxy, as indicated by the low fraction of gray pixels (in fact, a majority of them are either black or white), and by the low value of $R_{0.5} = 0.04$. Besides, with a $F_1 = 0.93$, this is also an excellent example to illustrate the capacity of SAM to retrieve reliable galaxy truncations and sizes.

The choice of fixing a given probability threshold (e.g., 0.5 for the majority voting criteria) stems from the aim of this study: to develop a methodology for inferring stellar disk truncations in galaxy images in an automated and consistent manner. The 0.5 threshold was chosen as it balances precision and recall (see \autoref{sec:performance}), providing robust segmentation maps while maintaining sensitivity to faint features in the galaxy outskirts. Alternative thresholds, such as 0.63, which maximizes the $F_1$ score for our dataset, are explored in \autoref{sec:performance} to assess their impact on galaxy size measurements. While higher thresholds yield more conservative (i.e., purer) segmentations, they often exclude significant portions of the truncations (i.e., less complete), particularly in galaxies with diffuse outer regions. Consequently, the 0.5 threshold was determined to be optimal for this study, ensuring a balance between capturing the full extent of the truncation and avoiding the inclusion of spurious features, both leading to a precise determination of the sizes of the galaxies in our dataset (see \autoref{sec:results}). Importantly, this threshold is not specifically tuned for this dataset but is instead a general and arbitrary choice, demonstrating its potential applicability to other datasets without requiring fine-tuning.

As we describe in \autoref{sec:sizes}, galaxies showing large values of $\Delta R_{0.5}$ would need to be inspected to check the validity of the sizes inferred or directly excluded from the analysis to avoid introducing non-accurate sizes. For those cases, defining the truncation as the contour encompassing the regions with an average value above 0.7 (or a different probability threshold) could lead to a better estimate of the truncation and its corresponding size with respect to BT24. For these reasons, we make publicly available the catalogs containing the parameters that characterize the truncations obtained with SAM and the averaged segmentation maps, allowing the user to use a different probability threshold.

%%%%%%%%%%%%%%%%%%%%%%%%%%%%%%%%%%%%%%%%%%%%%%%%%%%%%%%%%%%%%%%%%%%%%%%%%%%%%%%%%%%%%%%%%%%%%%%%%%%%
%%%%%%%%%%%%%%%%%%%%%%%%%%%%%%%%%%%%%%%%%%%%%%%%%%%%%%%%%%%%%%%%%%%%%%%%%%%%%%%%%%%%%%%%%%%%%%%%%%%%
%%%%%%%%%%%%%%%%%%%%%%%%%%%%%%%%%%%%%%%%%%%%%%%%%%%%%%%%%%%%%%%%%%%%%%%%%%%%%%%%%%%%%%%%%%%%%%%%%%%%

\section{Results}
\label{sec:results}

\begin{figure*}
\centering
	\includegraphics[width=\columnwidth]{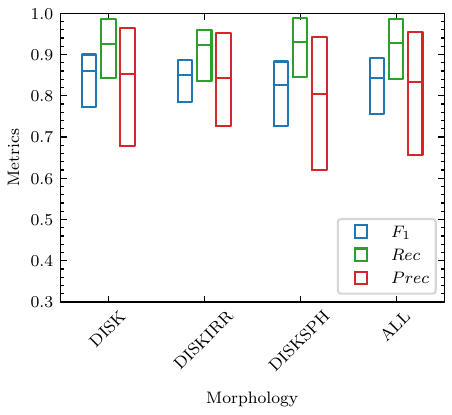}
	\includegraphics[width=\columnwidth]{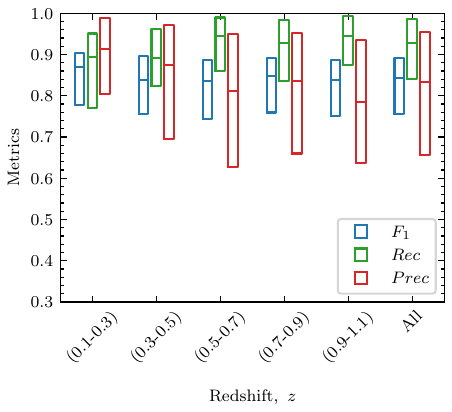}
	\includegraphics[width=\columnwidth]{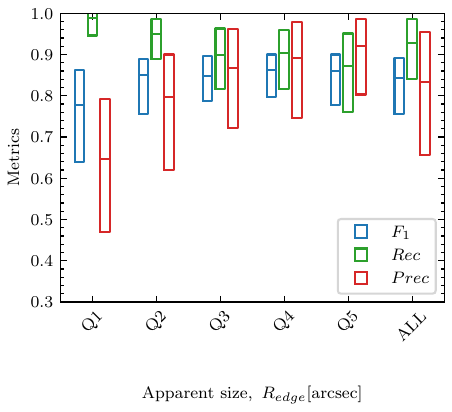}
	\includegraphics[width=\columnwidth]{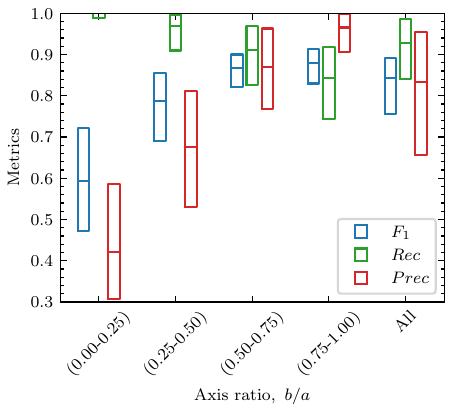}
	
    \caption{From left to right, SAM's performance as a function of: the morphological type, the redshift, the apparent size ($R_{edge}$, semi-major axis of the BT24 truncation) split in quintiles, and the axis ratio of the BT24 truncation. Metrics are shown in different colors: $F_1$ in blue, $Rec$ in green, and $Prec$ in red. Boxes represent the interquartile range (IQR), while the horizontal lines indicate the median value. $F_1$ and $Prec$ values decrease, while keeping high $Rec$ values, for small and/or elongated galaxies.}
    \label{fig:metrics_boxplot}
\end{figure*}

We aim to construct a model for determining stellar truncations in galaxy images, avoiding, to some extent, the need for specific preprocessing for each input image and survey. Additionally, we bypass the necessity of relying on a particular dataset with derived truncations (such as BT24) for training, fine-tuning, or evaluating the SAM predictions for truncations. This is one of the advantages of using an AI foundational model like SAM. It should be noted that retrieving truncations using traditional methods, such as the approach presented in BT24, requires significant effort. In addition, masking secondary sources in the images, as a preliminary step before computing the surface brightness and color profiles, is partially manual work that demands considerable expertise.

Hereafter, the truncations presented in BT24 are only used for comparison in the following sections.

The data supporting the findings of this study, including galaxy size measurements, averaged segmentation masks, and visualizations, are publicly available on GitHub\footnote{\url{https://github.com/astrovega/SAM-Euclid-sizes}}.

%%%%%%%%%%%%%%%%%%%%%%%%%%%%%%%%%%%%%%%%%%%%%%%%%%%%%%%%%%%%%%%%%%%%%%%%%%%%%%%%%%%%%%%%%%%%%%%%%%%%
%%%%%%%%%%%%%%%%%%%%%%%%%%%%%%%%%%%%%%%%%%%%%%%%%%%%%%%%%%%%%%%%%%%%%%%%%%%%%%%%%%%%%%%%%%%%%%%%%%%%

\subsection{SAM performance on retrieving truncations}
\label{sec:performance}

Beyond the results obtained for a particular galaxy, we compute the metrics introduced in \autoref{sec:sam_truncations} based on the truncations inferred with SAM and the estimates presented in BT24 for the whole dataset.

In \autoref{fig:metrics_boxplot}, we examine in detail these metrics as a function of the properties of the galaxies in the dataset: the apparent size of the BT24 truncation ($R_{edge}$, in arcsec), the redshift ($z$), the axis ratio ($b/a$), and the morphology. Overall, we obtain $F_1 = 0.84$, $Prec = 0.84$ and $Rec = 0.92$. This high value of $Rec$ reflects the fact that SAM successfully recovers a large fraction of the pixels within the truncation of BT24. Besides, the lower $Prec$ (equivalent to purity) compared to $Rec$ (or completeness) indicates that SAM truncations extend beyond those derived by BT24. The harmonic mean of these two quantities, the $F_1$ score, is still at an excellent level despite that SAM is only fed with RGB galaxy images and without the need for training nor fine-tuning the model (rather than surface brightness and mass profiles, and a labeled dataset, as done in BT24 and \citealt{Fernandez2024}). Metrics are relatively constant, with median values above 0.8, independently of the galaxy morphology and redshift. However, $F_1$ and $Prec$ values drop for galaxies in the low-end of the apparent size and axis ratio distributions. In both cases, $Rec$ values are almost equal to 1 values, indicating that SAM tends to recover more extended truncations than BT24. Also, $Rec$ values tend to decrease slightly towards the high-end of the $R_{edge}$ distribution. Since cutouts have a fixed size, the smallest galaxies have a lower number of \textit{positive} pixels. This means that a small number of FP can have a strong effect on the $Prec$. On the other hand, larger galaxies have a larger number of \textit{positive} pixels, a factor on the denominator in the $Prec$ definition, explaining the drop of this quantity with increasing size. Additionally, small and elongated galaxies are more affected by the PSF, which makes them look puffier and, consequently, leads SAM to infer slightly more extended truncations. This effect is more evident towards the minor axis of the truncation (see the two examples in \autoref{fig:small_axis_ratio}) and, thus, does not strongly affect the estimated size with SAM ($R_{0.5}$).

To reinforce the choice of the 0.5 threshold in the majority voting criteria, we computed the threshold that maximices the  $F1$ for our dataset when comparing with BT24. We found the median value of this optimal threshold to be 0.63, which results in $F_1 = 0.84$, $Prec = 0.89$, and $Rec = 0.88$. Differences in $F_1$ with respect to the 0.5 threshold for the majority voting criteria are negligible (on the third decimal). However, increasing the threshold value from 0.5 to 0.63 improves the purity of the retrieved truncations (i.e., higher $Prec$) at the expense of reduced completeness (i.e., lower $Rec$). In practical terms, this means retrieving less extended truncations —and consequently smaller galaxy sizes— when using a higher threshold, which are not in the same level of agreement as those obtained with the 0.5 threshold in the majority voting criteria.

The 11 galaxies with the largest truncations according to BT24 in our dataset present truncations that extend beyond the image limits of $12 \times 12$ arcsec. These galaxies are not included in the analysis presented hereafter. In these cases, SAM has more difficulty accurately retrieving the stellar disk truncations because the galaxy cannot be completely shown in the image and there is not enough background field around the main galaxy for the model to contrast with. In fact, for 6 of these galaxies, SAM tends to underestimate the size of the truncations compared to BT24. To avoid this, it would be necessary to produce larger images for the more extended galaxies and check if the truncations are better identified compared to the standard image sizes used for the whole dataset. While SAM can easily handle varying sizes and aspects of input images (i.e., they do not need to be of a fixed size), this is beyond the scope of the current study. We will consider producing images with a field of view that better accommodates the apparent size of the galaxy when applying this methodology to real Euclid data in the future.

%%%%%%%%%%%%%%%%%%%%%%%%%%%%%%%%%%%%%%%%%%%%%%%%%%%%%%%%%%%%%%%%%%%%%%%%%%%%%%%%%%%%%%%%%%%%%%%%%%%%
%%%%%%%%%%%%%%%%%%%%%%%%%%%%%%%%%%%%%%%%%%%%%%%%%%%%%%%%%%%%%%%%%%%%%%%%%%%%%%%%%%%%%%%%%%%%%%%%%%%%

\subsection{Galaxy sizes inferred from SAM truncations}
\label{sec:sizes}

\begin{figure*}
\centering
	\includegraphics[width=\columnwidth]{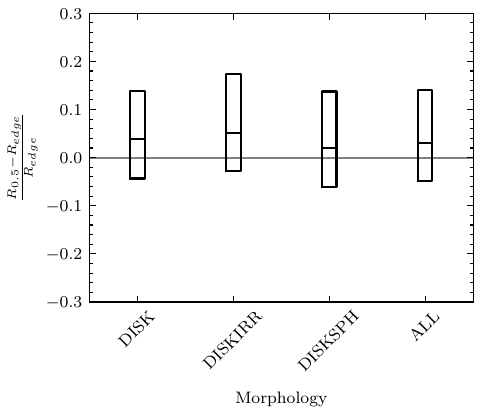}
	\includegraphics[width=\columnwidth]{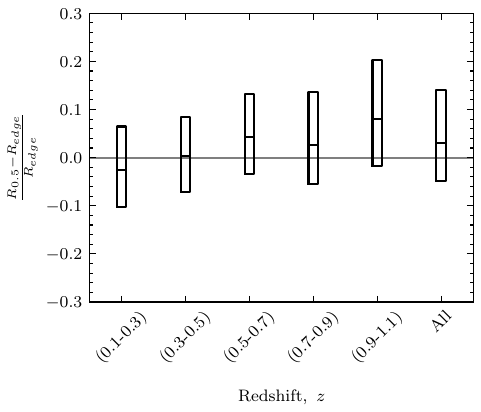}
	\includegraphics[width=\columnwidth]{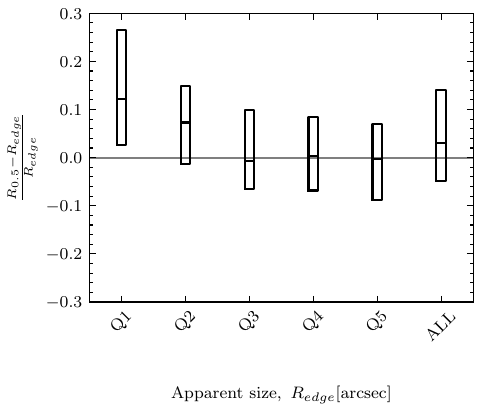}
	\includegraphics[width=\columnwidth]{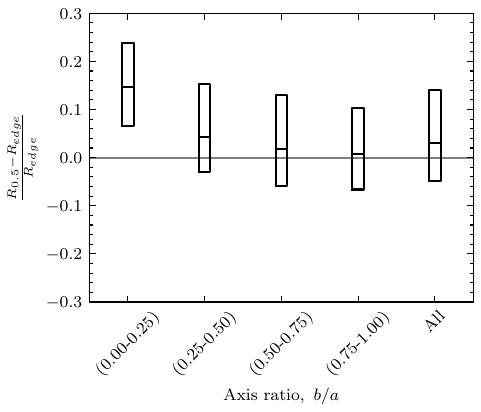}
	
    \caption{From left to right, relative difference between the size of the galaxy measured from the SAM truncation (denoted as $R_{0.5}$) and the semi-major axis of the truncation ellipse in BT24 (denoted as $R_{edge}$) as a function of: the morphological type, the redshift, the apparent size ($R_{edge}$, semi-major axis of the BT24 truncation) split in quintiles, and the axis ratio of the BT24 truncation. Boxes represent the interquartile range (IQR), while the horizontal lines indicate the median value. SAM truncations are on average overestimated with respect to BT24 truncations for the most elongated (in projection) and/or for the (apparently) smallest galaxies.}
    \label{fig:size_boxplots}
\end{figure*}

As described in \autoref{sec:sam_truncations}, we estimate the size of the galaxy, denoted as $R_{0.5}$, as the maximum distance from the center of the galaxy to the SAM truncation from the majority voting criteria. This choice of galaxy size, as measured in the averaged segmentation maps, is made to facilitate comparison with the size derived by BT24 as the semi-major axis of the elliptical truncation. We derive the relative error at measuring galactic sizes between the SAM predictions ($R_{0.5}$) and the BT24 estimates (i.e., the semi-major axis of the elliptical truncation, $R_{edge}$), as follow: $(R_{0.5}-R_{edge})/R_{edge}$. In \autoref{fig:size_boxplots}, we show the relative error in the size estimates as a function of different galaxy parameters. We find median size values inferred with SAM to be overestimated with respect to BT24, independently of the morphological type, although the median relative error is always $< 5 \%$. There is not a clear dependence with redshift, although the median values of the size relative error are below the $8\%$ for all the redshift bins analyzed. The largest discrepancies (between $12-14\%$ relative errors in size) are due to galaxies with the smallest apparent sizes (i.e., those within the Q1 of the distribution), and the most elongated galaxies (i.e., $b/a \lesssim 0.25$), for which SAM tends to overestimate the sizes of galaxies with respect to BT24. 

\begin{figure}
\centering
	\includegraphics[width=\columnwidth]{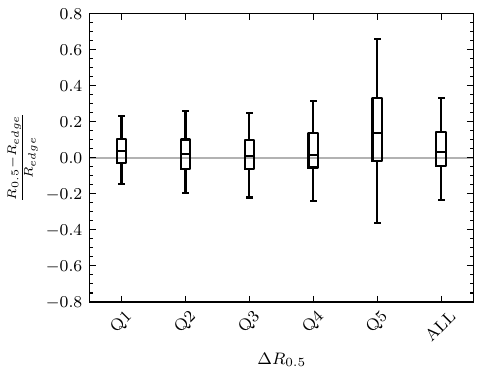}
 
    \caption{Relative error on the size of the truncation inferred with SAM with respect to the BT24 measurements as a function of the relative error on the size of the truncation derived from the averaged SAM segmentation maps ($\Delta R_{0.5}$) split in quintiles. Boxes represent the interquartile range (IQR) and whiskers extend from the box to the farthest data point lying within $1.5\times\mathrm{IQR}$ from the box, while the horizontal lines indicate the median value. Median values (and the scatter) of the relative error on the size obtained with SAM compared to BT24 are below the $\sim 5\%$ level for all the quintiles with the exception of Q5. For the Q5, where the differences between the size of the truncation at a 0.5 and a 0.7 probability thresholds are the largest, the median value of the relative error on the size of the truncation inferred with SAM surpasses the $\sim 10\%$ level with a clear increase on the scatter.}
    \label{fig:size_Rth_boxplot}
\end{figure}

Using the definition of the relative error in the measurement of the size of the SAM truncation described in \autoref{sec:sam_truncations}, denoted as $\Delta R_{0.5}$, we examine the robustness and reliability of the galactic sizes inferred with SAM. In \autoref{fig:size_Rth_boxplot}, we show the distribution of the relative error on the size of the truncation inferred with SAM with respect to the BT24 measurements, denoted as $(R_{0.5}-R_{edge})/R_{edge}$,  as a function of $\Delta R_{0.5}$. The median values of $(R_{0.5}-R_{edge})/R_{edge}$ are constrain within less than $\sim0.05$ and interquartile range (IQR) below $\sim0.10$ for the Q1-Q4 quintiles. For the largest values of the $\Delta R_{0.5}$ distribution (the Q5 quintile), median relative errors exceed the $\sim 0.10$ level. Besides, the most extreme values of $(R_{0.5}-R_{edge})/R_{edge}$ correspond to those cases with $\Delta R_{0.5}$ in the Q5 quintile, as shown by the length of the whiskers. Therefore, it is possible to identify cases with large relative errors in the sizes derived with SAM and with respect to BT24 estimates by selecting those with large $\Delta R_{0.5}$ values, without depending on the truncations and sizes derived by BT24. We illustrate this in \autoref{fig:high_deltaR}, where we show the SAM truncations for two galaxies with  $\Delta R_{0.5}$ in the Q5 quintile. 

\begin{figure}
\centering
	\includegraphics[width=\columnwidth]{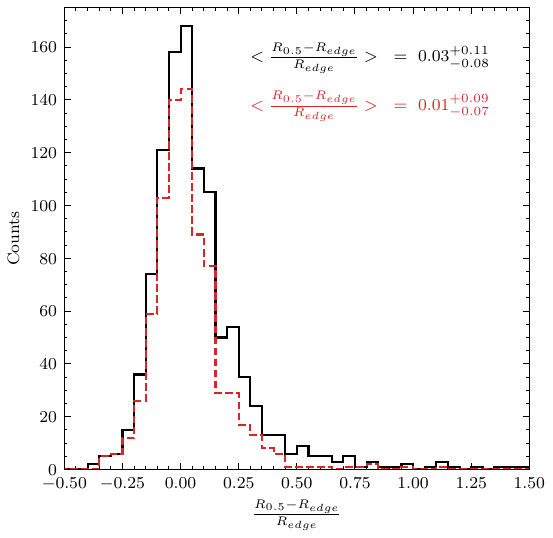}
 
    \caption{Distributions of the relative error between the size of the galaxy measured from the SAM truncation (denoted as $R_{0.5}$) and the semi-major axis of the truncation ellipse in BT24 (denoted as $R_{edge}$). The black solid histogram corresponds to the whole sample of 1047 galaxies, while the red dashed histogram shows the distribution for 774 galaxies with $\Delta R_{0.5} \lesssim 0.17$ (i.e., excluding those within the Q5 quintile) and $b/a$ > 0.25. The median and IQR of the relative error between the two estimates for both distributions are also shown.}
    \label{fig:hist_radius}
\end{figure}

Overall, in \autoref{fig:hist_radius}, we find a relative error for the size of the galaxy of $(R_{0.5}-R_{edge})/R_{edge}=0.03^{+0.11}_{-0.8}$ for the whole dataset, expressed as the median value and the IQR of the distribution. Then, we exclude from this analysis galaxies for which the sizes derived with SAM are less reliable: $\Delta R_{0.5} \gtrsim 0.17$ (those within the Q5 quintile, see \autoref{fig:size_Rth_boxplot}) and elongated in projection with $b/a$ < 0.25 (those within the Q1 quintile, see the bottom-right panel in \autoref{fig:size_boxplots}). It is clear how this selection excludes cases mostly on the high-end of the $(R_{0.5}-R_{edge})/R_{edge}$ distribution, which leads to a slightly better averaged size estimate, $(R_{0.5}-R_{edge})/R_{edge}=0.01^{+0.9}_{-0.7}$. In \autoref{fig:large_errorR_clean}, we show two examples of galaxies within the high-end of the $(R_{0.5}-R_{edge})/R_{edge}$ distribution, even though they fall within the Q1-Q4 quintiles of the $\Delta R_{0.5}$ distribution. However, it will be possible to identify galaxies with close companions, which are likely to have less reliable SAM truncations and sizes, using the source catalogs that will be published along with the Euclid images.

This excellent result demonstrates the enormous potential of using SAM for retrieving galaxy sizes in large datasets such as the Euclid Wide Survey in an automated way, without the need to train the model or a heavy preprocessing of the input images. Moreover, as can be seen in the right-hand panel in \autoref{fig:large_errorR_clean}, SAM can help retrieve truncations more accurately in non-symmetric galaxies that could be problematic for other approaches that rely on the assumption of elliptical symmetry for the truncations.

%%%%%%%%%%%%%%%%%%%%%%%%%%%%%%%%%%%%%%%%%%%%%%%%%%%%%%%%%%%%%%%%%%%%%%%%%%%%%%%%%%%%%%%%%%%%%%%%%%%%
%%%%%%%%%%%%%%%%%%%%%%%%%%%%%%%%%%%%%%%%%%%%%%%%%%%%%%%%%%%%%%%%%%%%%%%%%%%%%%%%%%%%%%%%%%%%%%%%%%%%
%%%%%%%%%%%%%%%%%%%%%%%%%%%%%%%%%%%%%%%%%%%%%%%%%%%%%%%%%%%%%%%%%%%%%%%%%%%%%%%%%%%%%%%%%%%%%%%%%%%%

\section{Conclusions}
\label{sec:conclusions}

In this study, we explore the potential of using an AI foundational model, specifically SAM, to retrieve galactic sizes in upcoming Euclid data from the stellar disk truncation. Our findings suggest that SAM can significantly optimize and accelerate the process of galactic size measurement in an automated manner. Although our current results are not yet competitive with those achieved by supervised models (e.g., \citealt{Fernandez2024} obtained an $F_1 = 0.91$), SAM eliminates the need for extensive training or fine-tuning typically required by these models, achieving significantly high performance. Furthermore, without relying on any physical assumptions about the shape of the truncation, SAM can retrieve reliable truncations and sizes for non-symmetric galaxies —cases that can be problematic when addressed by other approaches.

SAM's segmentation tasks were conducted on a system with an NVIDIA RTX A5000 GPU (24 GB RAM) and an AMD Ryzen 9 5900X 12-Core Processor. With this setup, SAM can infer galaxy sizes with a $\approx 3\%$ error ($\approx 1\%$ error if unreliable measurements are excluded) with respect to BT24, with an average runtime of approximately 1 minute per galaxy (i.e., roughly 2 seconds per configuration), considering all 24 different configurations per galaxy described in \autoref{sec:selection}. While GPU acceleration significantly reduced processing time, SAM can also operate on CPU systems, albeit with slower runtimes. The model's scalability is advantageous for processing large datasets like Euclid Wide Survey, where segmentation tasks could be parallelized across multiple GPUs or computational nodes.

Still, the entire process could be optimized further, for instance, by reducing the number of pixels (e.g., through re-binning) and the number of input configurations per galaxy. Moreover, different observational scenarios (e.g., less massive or low surface brightness galaxies) from those described in this study might require additional tuning during the preprocessing of input images. For example, for galaxies with low surface brightness or at higher redshifts than those in our dataset (where surface brightness dimming is significant), increasing the product $q \times s$ and the contrast ($c$) in the preprocessing stage can enhance the visibility of faint outer regions, leading to improved segmentation accuracy. Different optimizations will be explored when applying this methodology to the upcoming Euclid data, ensuring the robustness and scalability of SAM for larger and more diverse datasets.

Using the optimal input configurations presented in \autoref{sec:selection}, we demonstrated excellent performance in retrieving stellar disk truncations and galaxy sizes for the galaxies in our dataset (as shown in \autoref{sec:results}). It is worth noting that our dataset already encompasses a broad variety of sources in terms of galaxy sizes, axis ratios, and redshift coverage (up to $z \sim 1$). However, future applications of SAM to datasets of galaxies with significantly different properties, such as non-disk-like galaxies, may benefit from systematic parameter optimization.

Nonetheless, the results presented in this manuscript may serve as a proof of concept and represent a substantial advancement in the field, allowing researchers to bypass the time-consuming stages and human supervision required by other traditional methods, such as those described in \cite{Chamba2022} and \cite{Buitrago2024}.

During the preparation of this manuscript, the Meta AI\footnote{\url{https://ai.meta.com/}} team published an updated model for segmentation tasks, SAM 2 \citep{ravi2024sam2}. SAM 2 extends the previous version by performing visual segmentation in videos. We apply the whole pipeline described in this manuscript using SAM 2 over our dataset to retrieve stellar disk truncations and galactic sizes in the same way as done for SAM. Surprisingly, the results obtained with SAM 2 are not as satisfactory as those retrieved with SAM. After a closer examination of the results, we detect that SAM 2 tends to segment more frequently the inner parts of the galaxies in the dataset. This leads to a better purity ($Prec = 0.90$) but worse completeness ($Rec = 0.88$), although showing a similar median $F_1 = 0.84$. Consequently, the majority voting criteria applied to SAM 2 averaged segmentation maps tends to underestimate the size of the galaxies in our dataset compared to BT24. Although with a different configuration, SAM 2 could also be a valid approach to infer galaxy truncations and sizes, we leave these tasks for future work and rely on the excellent results provided by SAM and described throughout this study. 

By leveraging SAM, we can achieve rapid and reliable galactic size estimations, which is crucial for large-scale astronomical surveys like Euclid Wide Survey. The integration of AI foundational models like SAM into astronomical data analysis holds great promise. In particular, SAM will allow us to estimate truncations for millions of galaxies up to $z \sim 2$ when applied to upcoming Euclid Wide Survey, paving the way to understand the physical processes that shape galaxy evolution across cosmic time. Accurate galaxy size measurements derived from stellar disk truncations are critical for refining the mass-size relation, a fundamental aspect of understanding galaxy evolution. These measurements can also inform studies of galaxy morphology and star formation mechanisms, providing insight into how galaxies evolve in different cosmic environments. Within the context of Euclid, precise size estimates will contribute to mapping the distribution of galaxies across large-scale structures and understanding the interplay between galaxy evolution and dark matter halos, key to Euclid’s cosmological goals. 

\begin{acknowledgements}

JV-F acknowledges Ra\'ul Infante-Sainz and Mohammad Akhlaghi for their insightful comments, and advice, which significantly contributed to the development of this work. JV-F, FB, JF-I and SR acknowledge support from the project the GEELSBE project with reference PID2020-116188GA-I00, funded by MICIU/AEI /10.13039/501100011033. JV-F, FB, JF-I, SR, BS and HDS also acknowledge support from the GEELSBE2 project with reference PID2023-150393NB-I00 funded by MCIU/AEI/10.13039/501100011033 and the FSE+. JV-F and HDS acknowledge the financial support from the European Union - NextGenerationEU and the Spanish Ministry of Science and Innovation through the Recovery and Resilience Facility project ICTS-MRR-2021-03-CEFCA and the support from the grant PID2022-138896NA-C54. HDS acknowledges financial support by RyC2022-030469-I grant, funded by MCIN/AEI/10.13039/501100011033  and FSE+. This work has made use of the Rainbow Cosmological Surveys Database, which is operated by the Centro de Astrobiología (CAB/INTA), partnered with the University of California Observatories at Santa Cruz (UCO/Lick,UCSC). Based on zCOSMOS observations carried out using the Very Large Telescope at the ESO Paranal Observatory under Programme ID: LP175.A0839. We have used extensively the following software packages: MATPLOTLIB \citep{Hunter2007}, PYTORCH \citep{Paszke2019}, and Segment Anything Model \citep{Kirillov2023}.

\end{acknowledgements}

\bibliographystyle{aa}
\bibliography{bibliography} % if your bibtex file is called example.bib

\begin{appendix}

\section{Examples of galaxy truncations derived with SAM}

In this appendix, we show a selection of galaxy images to illustrate the potential of SAM to derive stellar disk truncations in `euclidized' galaxy images. All the images shown in this section are produced with the Gnuastro script \texttt{astscript-color-faint-gray} for $q \times s = 100$ and $c = 1$.

\begin{figure}
\centering
	\includegraphics[width=\columnwidth]{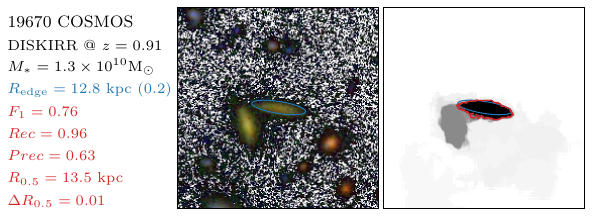}
	\includegraphics[width=\columnwidth]{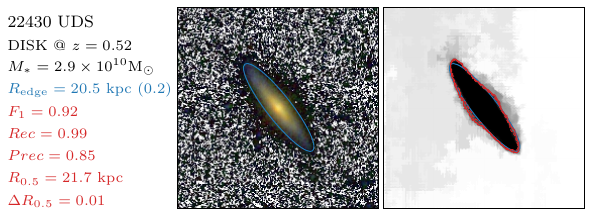}
	
    \caption{Galaxy images with elongated shapes ($b/a < 0.25$). Image details are described in the caption of \autoref{fig:sam_truncations}.}
    \label{fig:small_axis_ratio}
\end{figure}

In \autoref{fig:small_axis_ratio}, we show two galaxies with extremely elongated shapes, i.e., $b/a < 0.25$, for which the $Prec$ and $F_1$ tend to be worse than for more rounded galaxies, as shown in \autoref{fig:metrics_boxplot}. The example in the left panel (19670 COSMOS) shows a galaxy with $b/a = 0.2$ and close to the high-end of the redshift distribution ($z = 0.91$). Although the central galaxy is close to another galaxy with apparently similar brightness, the SAM truncation and its estimated size are in good agreement with the findings of BT24. Nevertheless, we find a $F_1 = 0.76$ because of the SAM truncation extends beyond the BT24 truncation ($Prec = 0.63)$. The relative error in the size with respect to BT24 is $(R_{0.5}-R_{edge})/R_{edge} \approx 0.05$. The galaxy in the right panel (22430 UDS) has a $b/a = 0.2$ at a $z=0.53$ and, therefore, with a larger apparent size compared to the previous example. The SAM truncation is in excellent agreement with the findings by BT24, showing a $F_1 = 0.92$ and a relative error in the size of $(R_{0.5}-R_{edge})/R_{edge} \approx 0.06$. In both cases, the different estimates of the truncation are in good agreement for the different input configurations, as the low values of $\Delta R_{0.5} = 0.01$ reflect.

\begin{figure}
\centering
	\includegraphics[width=\columnwidth]{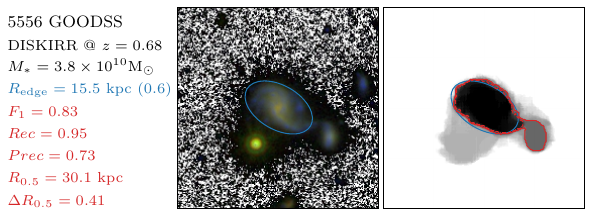}
	\includegraphics[width=\columnwidth]{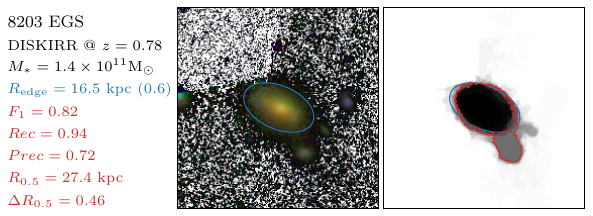}
	
    \caption{Galaxy images with $\Delta R_{0.5} \gtrsim 0.17$ (within the Q5 quintile). Image details are described in the caption of \autoref{fig:sam_truncations}.}
    \label{fig:high_deltaR}
\end{figure}

Examples of galaxy images, 5556 GOODSS and 8203 EGS, with large values of $\Delta R_{0.5} \gtrsim 0.17$ (within the Q5 quintile, see \autoref{fig:size_Rth_boxplot}) are shown in \autoref{fig:high_deltaR}. In both cases, the large values of $\Delta R_{0.5}$ are due to the presence of a neighbor galaxy close to the central galaxy. While the SAM truncation at a 0.5 probability threshold includes the close galaxy within the truncation, the contour at a 0.7 probability threshold leads to a better estimate of the truncation and its size. In these cases, it is possible to correct the size of the SAM truncation by the factor $\Delta R_{0.5}$ to better agree with the results obtained by BT24. By doing so, also the metrics shown in the figure would increase.  

\begin{figure}
\centering
	\includegraphics[width=\columnwidth]{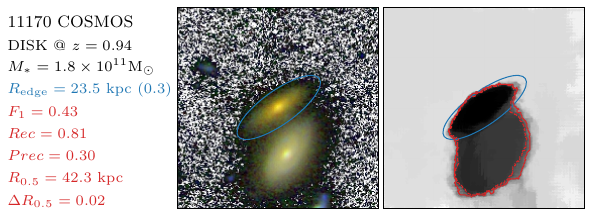}
	\includegraphics[width=\columnwidth]{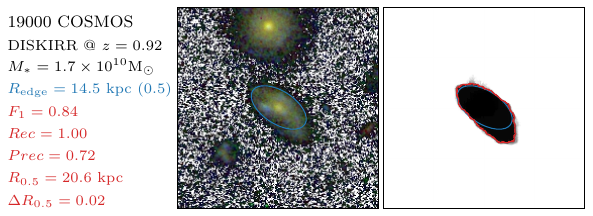}
	
    \caption{Galaxy images with $\Delta R_{0.5} \lesssim 0.17$ (within the Q1-Q4 quintile) and $(R_{0.5}-R_{edge})/R_{edge} > 0.20$. Image details are described in the caption of \autoref{fig:sam_truncations}.}
    \label{fig:large_errorR_clean}
\end{figure}

In \autoref{fig:large_errorR_clean}, we show two examples of galaxies, 11170 COSMOS and 19000 COSMOS, for which the size estimate with SAM of more than $20\%$ larger than the size measured by BT24, even though they both exhibit small values of $\Delta R_{0.5} = 0.02$. These cases correspond to galaxies at the high-end of the $(R_{0.5}-R_{edge})/R_{edge}$ distribution shown in \autoref{fig:hist_radius} when $\Delta R_{0.5} \gtrsim 0.17$ are not included, and for which neither the truncation defined at 0.5 or 0.7 probability thresholds lead to a good estimate of the truncation and its size. Therefore, to obtain a better estimate of the size of the truncation for 11170 COSMOS with respect to BT24, the truncation should be derived for an even higher probability threshold. For 19000 COSMOS, the difference between $R_{0.5}$ and $R_{edge}$ is due to a bright extension of the galaxy toward the bottom-right corner of the figure. This structure, likely a tidal feature produced by a past interaction of the central galaxy with another galaxy, seems to be part of the central galaxy. Therefore, the measured size of the truncation with SAM might not be as inaccurate as it is when compared with the truncation derived by BT24.

%%%%%%%%%%%%%%%%%%%%%%%%%%%%%%%%%%%%%%%%%%%%%%%%%%%%%%%%%%%%%%%%%%%%%%%%%%%%%%%%%%%%%%%%%%%%%%%%%%%%
%%%%%%%%%%%%%%%%%%%%%%%%%%%%%%%%%%%%%%%%%%%%%%%%%%%%%%%%%%%%%%%%%%%%%%%%%%%%%%%%%%%%%%%%%%%%%%%%%%%%
%%%%%%%%%%%%%%%%%%%%%%%%%%%%%%%%%%%%%%%%%%%%%%%%%%%%%%%%%%%%%%%%%%%%%%%%%%%%%%%%%%%%%%%%%%%%%%%%%%%%

\end{appendix}
\end{document}